\documentclass[5p,times]{elsarticle}
\usepackage{hyperref}

\usepackage{multirow}
\usepackage{graphicx}
\usepackage{stmaryrd}
\usepackage{footmisc}
\usepackage{amssymb}

\usepackage{lineno}
\usepackage{algorithm}
\usepackage{algpseudocode}
\usepackage[font=footnotesize,labelfont=bf]{caption}
\usepackage{subcaption}
\usepackage{url}
\usepackage{booktabs}

\usepackage{xspace}
\newcommand{\ie}{\textit{i.e.}\xspace}
\newcommand{\eg}{\textit{e.g.}\xspace}

\newcommand{\cf}{\textit{cf.}\xspace}

\journal{Information and Software Technology}

\begin{document}
\makeatletter
\def\ps@pprintTitle{%
	\let\@oddhead\@empty
	\let\@evenhead\@empty
	\def\@oddfoot{\parbox[t]{1\linewidth}{\footnotesize \textit{Accepted Manuscript. \@journal. \\ Available online at ScienceDirect since 1 November 2022. \\ ~\url{https://doi.org/10.1016/j.infsof.2022.107100} \\
				License: \href{https://creativecommons.org/licenses/by-nc-nd/4.0/}{CC BY-NC-ND 4.0} }}}%
	\let\@evenfoot\@oddfoot}
\makeatother

\begin{frontmatter}

\title{A Property Specification Pattern Catalog for Real-Time \\ System Verification with UPPAAL}

\author[hum]{Thomas Vogel\corref{cor1}}
\author[hum]{Marc Carwehl}
\author[cic]{Gena\'ina Nunes Rodrigues}
\author[hum]{Lars Grunske}

\address[hum]{Institut für Informatik, Humboldt-Universität zu Berlin, Germany \\
\{thomas.vogel,carwehl,grunske\}@informatik.hu.berlin.de}
\address[cic]{Department of Computer Science, University of Bras\'ilia, Brazil \\ genaina@unb.br}

\begin{abstract}
\textbf{Context:} The goal of specification pattern catalogs for real-time requirements is to mask the complexity of specifying such requirements in a timed temporal logic for verification. For this purpose, they provide frontends to express and translate pattern-based natural language requirements to formulae in a suitable logic. However, the widely used real-time model checking tool UPPAAL only supports a restricted subset of those formulae that focus only on basic and non-nested reachability, safety, and liveness properties. This restriction renders many specification patterns inapplicable. As a workaround, timed observer automata need to be constructed manually to express sophisticated requirements envisioned by these patterns. 
\textbf{Objective:} In this work, we fill these gaps by providing a comprehensive specification pattern catalog for UPPAAL. The catalog supports qualitative and real-time requirements and covers all corresponding patterns of existing catalogs.
\textbf{Method:} The catalog we propose is integrated with UPPAAL. It supports the specification of qualitative and real-time requirements using patterns and provides an automated generator that translates these requirements to observer automata and TCTL formulae. The resulting artifacts are used for verifying systems in UPPAAL. Thus, our catalog enables an automated end-to-end verification process for UPPAAL based on property specification patterns and observer automata.
\textbf{Results:} We evaluate our catalog on three UPPAAL system models reported in the literature and mostly applied in an industrial setting. As a result, not only the reproducibility of the related UPPAAL models was possible, but also the validation of an automated, seamless, and accurate pattern- and observer-based verification process.
\textbf{Conclusion:} The proposed property specification pattern catalog for UPPAAL enables practitioners to specify qualitative and real-time requirements in a pattern-based way---without directly using a temporal logic---and to verify them in UPPAAL while supporting a comprehensive set of patterns. 
\end{abstract}


\end{frontmatter}

\sloppy
\section{Introduction}\label{sec:Introduction}

Since the fundamental results by Alur et al.~\cite{Alur1996,Alur91} on decidability of model checking for real-time systems, a number of tools for automatic verification of hybrid and real–time systems have emerged~\cite{KronosCAV98,bllpw:dimacs95, KNP11}. Among those, the model checker UPPAAL~\cite{bllpw:dimacs95} has been applied successfully to several case studies~\cite{UPPAALCS21} to verify reachability, safety, and liveness properties of real-time systems~\cite{UPPAALtutorial}. 
Nevertheless, the formal specification of real-time requirements using a timed temporal logic is still considered an error-prone and time-consuming task (cf.~\cite{BelliniNR09}). Often practitioners are not familiar with the particular logic and regard it as too difficult to use a timed temporal logic to specify requirements. An example of a timed temporal logic is Timed Computation Tree Logic~(TCTL)~\cite{HENZINGER1994193,AlurCD90}, a subset of which is supported by UPPAAL~\cite{UPPAALtutorial}.

\textit{Property specification patterns} have been used to bridge this gap between practitioners and model checking~\cite{DwyerAC99, Gruhn06L}. Such patterns provide ``{general rules that help practitioners to qualify order and occurrence, to quantify time bounds, and to express probabilities of events}''~\cite[p.\,620]{AutiliGLPT15} when specifying properties to reason about events in reactive systems. As typical for patterns, the property specification patterns are organized in a catalog~\cite{Gruhn06L, AutiliGLPT15, KonradCheng05, Grunske08}. According to~\citet{DwyerAC99}, a property specification pattern catalog comprises the ``best practices'' in system specification and represents an attempt to capture proven solutions in a single framework. To align different formalities behind various specification pattern catalogs and the required knowledge to formalize system properties by non-experts, \citet{AutiliGLPT15} have proposed a comprehensive pattern catalog that unifies qualitative~\cite{DwyerAC99}, real-time~\cite{KonradCheng05}, and probabilistic~\cite{Grunske08} properties in a structured natural language interface. Each pattern of this catalog is expressed in a structured natural language and mapped to compatible temporal logic formalisms. This mapping enables an automated translation of properties expressed in the structured natural language to a temporal logic, which eases the formal specification of requirements. The unified catalog comprises 58 patterns exploiting the full expressive power of temporal logic formalisms.

From a theoretical perspective, the expressive power of a temporal logic is high, for instance, because modal operators can be nested to express sophisticated requirements and extended by time bounds to express real-time requirements. However, from a practical perspective, existing model checkers often support only a subset of a temporal logic to realize an efficient model checking. Particularly, UPPAAL does not support the nesting and timed extensions of modal operators. Thus, only a subset of TCTL properties can be expressed and verified by UPPAAL, especially basic reachability, safety, and liveness properties where timing aspects are restricted to clock constraints in state formulae~\cite{UPPAALtutorial,HavelundLS99}.
Consequently, there exists a gap between the property specification pattern catalog~\cite{AutiliGLPT15} that uses the full expressive power of temporal logic formalisms and the popular model checker UPPAAL that supports only a subset of such formalisms. 
Notably, the pattern catalog contains 18 patterns for qualitative and 19 patterns for real-time properties, each in five variants because of five different scopes, which results in a total of 185 pattern variants. Out of these 185 variants, only four variants can be directly expressed with UPPAAL, which focuses on (but is not limited to) verifying real-time systems.
Thus, UPPAAL does not support the majority of the property specification patterns.

To mitigate the limited support of TCTL by UPPAAL, Havelund et al. \cite{HavelundLS99} use a workaround. They sketch \textit{manual} techniques to annotate the system model and to define \textit{observer automata}, which have to be concretized for the property at hand. Particularly, new variables or communication actions are added to the system model, which are either referenced in the temporal logic formula to be verified or synchronized with an added observer automaton that monitors the system model during verification. For the latter, the observer reaches a certain state if the property is violated in the system model, which is checked by a new formula to be verified that refers to the corresponding state of the observer. Thus, the added variables or the observer encode those aspects that cannot be expressed by the subset of TCTL supported by UPPAAL. 
However, \citet{HavelundLS99} do not automate and connect these techniques to property specification patterns.

In this context, only a few approaches exist. 
\citet{Gruhn06L} recognized the gap between real-time property specification patterns and existing model checkers, and presented a catalog of such patterns each mapped to observer templates. However, this catalog does not support all real-time patterns collected by \citet{AutiliGLPT15} and it is not connected to a timed model checker by an automated generation of observers from the proposed templates.
\citet{Andre13} also proposes observer templates for real-time patterns that are a subset of the real-time patterns collected by \citet{AutiliGLPT15}, although with an automated generation of observers for the IMITATOR verification tool. Afterwards, \citet{AndreP15} specify these patterns in a dedicated language and define their semantics by Time Petri Nets, however, without extending the set of real-time property patterns.
Consequently, existing work does not bridge the gap between \textit{all} property specification patterns and existing model checkers considering real-time requirements by leveraging an automated approach to generate observers. Thus, a reconciliation between practitioners' expectations and real-time model checking tools is still required.

To close this gap, we propose a comprehensive property specification pattern catalog for UPPAAL. The catalog supports qualitative and real-time properties that are specified using patterns and provides a generator that automatically translates the specified properties to observer automata and UPPAAL's restricted TCTL formulae. To achieve this translation, we have concretized and automated the manual techniques by \citet{HavelundLS99} for each pattern of our catalog. Thus, our catalog leverages the benefits of property specification patterns to the widely-used UPPAAL by enabling practitioners to specify properties in a pattern-based way, that is, without using a temporal logic, \textit{and} to verify these properties in UPPAAL. 
Particularly, we make the following contributions in this paper:
(1)~As a theoretical contribution, our catalog comprises observer templates for \textit{all} specification patterns for qualitative and real-time properties collected by \citet{AutiliGLPT15}. The catalog is publicly available.\footnote{\url{https://github.com/hub-se/PSP-UPPAAL/wiki}}
(2)~As a practical contribution, our catalog is integrated with UPPAAL~\cite{bllpw:dimacs95}. It comes along with a generator that leverages the observer templates to automatically translate any property based on a pattern of our catalog to an observer automaton and TCTL formula used for the verification of systems modeled as timed automata in UPPAAL. Thus, the catalog and the generator enable an end-to-end verification process based on property specification patterns and observer automata.
(3)~We evaluate our catalog and generator on three real-time systems from literature~\cite{LindahlPY01,HavelundSLL97, RodriguesCR0P18}, two of which have been applied in an industrial setting. These systems have been originally verified with UPPAAL using manually created observer automata. Applying our catalog and generator to these systems, we demonstrate that we can express the properties of interest and generate automatically corresponding observer automata and formulae. Using these generated observer automata and formulae, we were able to reproduce the verification results reported in literature for these systems, which provides evidence for the validity of our catalog and generator.

The rest of the paper is organized as follows. 
We introduce a running example in Section~\ref{sec:BSN} and the theoretical background in Section~\ref{sec:Background}. 
In Section~\ref{sec:Proposal} we discuss our catalog and its methodology to automatically build observer automata based on property specification patterns. 
We evaluate our catalog on three real-time systems in Section~\ref{sec:Evaluation} and discuss related work in Section~\ref{sec:Related}.
Finally, we conclude and point out future work directions.

\section{Running Example: Body Sensor Network}\label{sec:BSN}

\begin{table*}[t]
	\centering
	\caption{Requirements of the Body Sensor Network (BSN).}
	\label{tab:bsn-properties}
	{\footnotesize
		\begin{tabular}{lp{0.85\textwidth}}
			\toprule
			\textbf{ID} &\textbf{Requirement} \\ \midrule
			BSN-P01 & The BSN is deadlock free. \\
			BSN-P02 & If a scheduler cycle is started, the BodyHub will be executed before the scheduler cycle ends. \\
			BSN-P03 & If a scheduler cycle is started, the three sensors will be executed before the scheduler cycle ends. \\
			BSN-P04 & If a sensor reports a health status of high risk, an emergency will be detected before one scheduler cycle elapses. \\
			BSN-P05 & If a sensor reports a health status of high risk, an emergency will be detected within 250 ms. \\
			BSN-P06 & If the patient's health status is on high risk, the corresponding number of scheduler cycles for high emergency risk status has not been exceeded.\\
			BSN-P07 & If the patient's health status is on moderate risk, the corresponding number of scheduler cycles for moderate emergency risk status has not been exceeded.\\
			BSN-P08 & If the patient's health status is on low risk, the corresponding number of scheduler cycles for low emergency risk status has not been exceeded.\\
			BSN-P09 & If data has been collected by the sensor node, the BodyHub will eventually process it.\\
			BSN-P10 & If data has been collected by the sensor node, the BodyHub will eventually persist it.\\
			BSN-P11 & If data has been sent by the sensor node, the BodyHub is able to process it as low, moderate or high risk vital sign data.\\
			BSN-P12 & If the BodyHub has processed data, it will eventually detect the patient's new health status.\\
			
			\bottomrule
		\end{tabular}
	}
\end{table*}

To illustrate our approach, we use the Body Sensor Network~(BSN)~\cite{RodriguesCR0P18} as a running example throughout this paper. The main objective of the BSN is to keep track of a patient's vital signs through three sensors and classify the patient's health status into \emph{low}, \emph{moderate}, or \emph{high} risk. In the case of high risk, the BSN sends an emergency signal to a central unit.
For this purpose, an electrocardiogram sensor, an oximeter, and a temperature sensor are connected to a patient and send their measurements to a central node (BodyHub) that persists and analyzes the data to determine the overall health status based on given thresholds for the measurements. 
Moreover, each sensor controls its sampling rate based on the health status locally determined from the sensor's own measurements. The sampling rate of a sensor effects also the rate of sending measurements to the BodyHub.

The BSN is further equipped with a scheduler that realizes the deterministic execution of the sensors and BodyHub using a first-come first-served (FCFS) strategy. 
Thus, the scheduler exclusively commands a sensor or the BodyHub to execute by sending fixed-period release signals. 

\citet{RodriguesCR0P18} have modeled the sensors, BodyHub and scheduler of the BSN as a network of timed automata in UPPAAL\footnote{\url{https://github.com/rdinizcal/SEAMS18/tree/master/uppaal}} and verified the requirements listed in Table~\ref{tab:bsn-properties}.
To enable verification, the BSN model contains a manually created observer.\footnote{We modified the original requirements to exclude any reference to the manually created observer used in the BSN model~\cite{RodriguesCR0P18}, and refined two requirements, each into two requirements for simplification.} 
The first requirement states that the BSN should be deadlock free (BSN-P01). 
BSN-P02 and BSN-P03 relate to fairness that all modules (the BodyHub and three sensors) will be executed in a scheduling cycle.
BSN-P04 and BSN-P05 specify that an emergency should be detected within 250 ms in a single scheduler cycle if the patient's health status is on high risk.
BSN-P06, BSN-P07, and BSN-P08 prescribe the behavior with respect to the sensor frequency at which data will be collected from the patient depending on the patient's health status (e.g., a more severe status implies a shorter sensing frequency). 
BSN-P09 and BSN-P10 state that the information collected from the sensor node is processed and respectively persisted by the BodyHub. 
Moreover, BSN-P11 prescribes that the sensor data sent to the BodyHub is in a certain range so that the BodyHub is able to classify the patient's health status as low, moderate or high risk. 
Finally, the requirement BSN-P12 prescribes that if the BodyHub has processed the collected data, it will detect the patient's latest health status.

\section{Background}\label{sec:Background}

In this section, we discuss the background of our work: 
UPPAAL and its limited support of full TCTL,
property specification patterns~\cite{AutiliGLPT15}, and 
manual techniques to cope with the limited support of TCTL by UPPAAL~\cite{HavelundLS99}.

\subsection{TCTL and its limited support by UPPAAL}
\label{sec:uppaal-tctl}

Timed Computation Tree Logic~(TCTL) is a variant of the Computation Tree Logic~(CTL) to express real-time properties~\cite{AlurCD90}. Such properties are specified and verified for systems modeled by timed automata~\cite{BaierPMC:2008}.
Like CTL properties, TCTL properties comprise state and path formulae. 
A state formula is an expression over the atomic propositions of a model and can be evaluated for each state of a system. 
A path formula quantifies over the paths of a model by referring to some path (\textit{E}) or to all paths (\textit{A}) and uses temporal operators to determine the states, for which the state formula should hold. Typical operators are \textit{G} for globally (\ie, now and forever in the future) and \textit{F} for finally (\ie, eventually in the future).
To express real-time properties, TCTL extends CTL with clock constraints over the clocks in the timed automaton that can be used in state formulae. Moreover, it extends CTL with time intervals for temporal operators to narrow the operator to a specified time window, for instance, $G^{[l,u]}$ and $F^{[l,u]}$ where $l$ and $u$ are the lower and upper bounds of the time interval. 

Independent of the TCTL extensions, typical properties that can be formulated are 
safety properties of the form $AG\varphi$ and $EG\varphi$,
and liveness properties of the form $AF\varphi$ and $EF\varphi$, where $\varphi$ is a state formula. To specify elaborate properties, the operators together with the path quantifiers can be nested. Examples are $AGAF\varphi$ and $AG(\varphi \rightarrow AF\psi)$, where $\varphi$ and $\psi$ are state formulae and $\rightarrow$ denotes the logical implication.
Additionally, all such formulae can use the TCTL extensions to express real-time properties.
To verify a property for a system, a model checker checks for a given timed automaton TA and TCTL formula $\phi$ whether TA satisfies $\phi$ (\ie, TA $\vDash$ $\phi$)~\cite{BaierPMC:2008}. 

\begin{table*}[htb]
    \caption{Types of TCTL formulae supported by UPPAAL~\cite{UPPAALtutorial}.}
	\label{tab:TCTL-UPPAAL}
    \centering
    \footnotesize
    \begin{tabular}{llp{0.7\textwidth}}
        \toprule
        \begin{tabular}{@{}l@{}}\textbf{TCTL} \\ \textbf{formula}\end{tabular} & \begin{tabular}{@{}l@{}}\textbf{UPPAAL} \\ \textbf{formula}\end{tabular} & \textbf{Description} \\ \midrule
        $AG\ \varphi$ & $\texttt{A[]}\ \varphi$ & $\varphi$ should be true in all reachable states, \ie, for all paths $\varphi$ is always true. \\
        $EG\ \varphi$ & $\texttt{E[]}\ \varphi$ & There should exist a maximal path for which $\varphi$ is always true, \ie, in every state of this path. \\
        $AF\ \varphi$ & $\texttt{A<>}\ \varphi$ & For all paths, $\varphi$ should be eventually true. \\
        $EF\ \varphi$ & $\texttt{E<>}\ \varphi$ & There should exist at least one path, for which $\varphi$ is eventually true. \\
        $AG(\varphi \rightarrow AF\ \psi)$ & $\varphi\ \texttt{-->}\ \psi$ & For all reachable states, whenever $\varphi$ is true, then eventually $\psi$ will be true. \\ 
        \bottomrule
    \end{tabular}
    \normalsize
\end{table*}

A popular model checker to verify TCTL properties on real-time systems is UPPAAL~\cite{bllpw:dimacs95}. A system is modeled as a network of timed automata that synchronize over channels, and properties are specified in a query language that supports a subset of TCTL~\cite{UPPAALtutorial}. 
Table~\ref{tab:TCTL-UPPAAL} lists the types of TCTL properties supported by UPPAAL. It can be seen that UPPAAL does \textit{not} support the nesting of path quantifiers and temporal operators (\ie, path formulae), except of the so-called response property shown in the last row of Table~\ref{tab:TCTL-UPPAAL}.
Concerning the real-time extensions of TCTL, UPPAAL does \textit{not} support timed temporal operators (\eg, $G^{[l,u]}$ and $F^{[l,u]}$), restricting the specification of timing aspects to clock constraints in state formulae.

Thus, UPPAAL's query language to specify properties is less expressive than TCTL, and not every TCTL formula can be expressed in UPPAAL.

\subsection{Property Specification Patterns}
\label{sec:property-patterns}

The goal of \textit{property specification patterns} is to ease the formalization of properties in a temporal logic for practitioners by capturing the knowledge of experts in formal methods~\cite{DwyerAC99}. The first patterns were proposed by \citet{DwyerAC99} for \textit{qualitative properties} and the occurrence and order of states or events. An \emph{occurrence} pattern describes that a state/event should or should not occur, while an \emph{order} pattern describes the relative sequence in which multiple states/events should occur during system execution. Each pattern has an intent and mappings to templates in different temporal logics such as CTL.\footnote{\url{https://matthewbdwyer.github.io/psp/}} 
Thus, practitioners select patterns based on the intents of the properties they want to formalize, and use the corresponding templates to create formal specifications.

Additionally, \citet{KonradCheng05} proposed new patterns for \textit{real-time properties}. Besides an intent for each pattern, they provide a specification in Structured English that is defined by a grammar as well as mappings to templates in different real-time temporal logics such as TCTL. Practitioners can use the Structured English grammar to express a property, and the corresponding mapping to obtain a formal specification of the property. 
Moreover, \citet{BelliniNR09} extended the patterns for real-time properties by providing real-time extensions for the qualitative properties by \citet{DwyerAC99}.

In addition to the qualitative and real-time patterns, \citet{Grunske08} proposed patterns for \textit{probabilistic properties} expressed in a Structured English grammar and mapped to templates in Continuous Stochastic Logic~(CSL).

Unifying these previous patterns~\cite{BelliniNR09,DwyerAC99,KonradCheng05,Grunske08} and proposing new ones, \citet{AutiliGLPT15} developed a comprehensive pattern catalog for qualitative, real-time, and probabilistic properties. The catalog provides a natural language interface in Structured English and mappings of each pattern to suitable temporal logics such as CTL, TCTL, and CSL.\footnote{\url{http://ps-patterns.wikidot.com}}
Since the focus of this paper is on qualitative and real-time properties, we consider the corresponding patterns while neglecting the patterns for probabilistic properties. 

As proposed by \citet{DwyerAC99}, each of these patterns have a \textit{scope} to define the fraction of the system execution for which the property must hold. 
Five different scopes exist: 
(1)~\textit{Globally}: the entire system execution,
(2)~\textit{Before~R}: the execution up to the first occurrence of the state or event~\textit{R}, 
(3)~\textit{After~Q}: the execution after the first occurrence of the state or event~\textit{Q},
(4)~\textit{Between~Q and~R}: any fragment of the execution beginning with the occurrence of state or event~\textit{Q} and ending with the occurrence of state or event~\textit{R},
and 
(5)~\textit{After~Q until~R}: any fragment of the execution as defined by the \textit{Between} scope, but the state or event \textit{R} does not need to~occur.\footnote{Similar to \citet{DwyerAC99}, we ignore nested fragments of $Q$ and $R$.}

\paragraph{Example}
To exemplify the formalization of a property with a pattern, we use the property BSN-P05 of the Body Sensor Network from Table~\ref{tab:bsn-properties}: 

\begin{quote}
    ``If a sensor reports a health status of high risk, an emergency will be detected within 250 ms.''
\end{quote}

A suitable pattern to formalize this property is the \textit{Response}, in the real-time variant because of the time constraint of detecting the emergency within 250 ms, and with the scope \textit{Globally} because the extent of the system execution is not limited.
For this pattern, \citet{AutiliGLPT15} provide a template in Structured English to specify a \textit{Response} property as well as a mapping to a TCTL template formalizing the property:

\begin{quote}
Structured English template:
\textbf{Globally}, \textbf{if} \textit{\{P\}} \textbf{[has occurred]} \textbf{then in response} \textit{\{S\}} \textbf{[eventually holds]} \textbf{within} \textit{$t_u^S$} \textbf{ms}. 

TCTL template: 
$AG(P \rightarrow AF^{[0,t_u^S]}(S))$
\end{quote}

The time constraint of the real-time response pattern is defined by $t_u^S$, which is the upper time bound for \textit{S} to occur after \textit{P} has occurred.\footnote{Besides an upper time bound resulting in a time interval $[0,t_u^S]$, \citet{AutiliGLPT15} further consider a lower time bound resulting in an interval $[t_l^S,\infty)$ as well as a combination to a time interval $[t_l^S,t_u^S]$ for the occurrences of states or events in all real-time patterns.}
Filling the placeholders \textit{P}, \textit{S}, and $t_u^S$ of the template for the given property we obtain:

\begin{quote}
\textbf{Globally}, \textbf{if} \textit{\{a sensor reports a health status of high risk\}} \textbf{then in response} \textit{\{an emergency will be detected\}} \textbf{within} \textit{250} \textbf{ms}. 
\end{quote}

Based on the mapping of the Structured English template to the TCTL template, we automatically translate the specification to a TCTL formula:

\begin{quote}
$AG(\{ \textrm{a sensor reports a health status of high risk} \} \rightarrow \hspace*{3em} AF^{[0,250]}(\{\textrm{an emergency will be detected}\}))$
\end{quote}

Thus, patterns support users in formally specifying properties by leveraging a natural language interface and automatically translating specification in Structured English to a temporal logic such as TCTL. 

However, the resulting TCTL formula of the example cannot be expressed in UPPAAL without introducing auxiliary formula clocks in the model. Though UPPAAL supports the (qualitative) response property (see last entry in Table~\ref{tab:TCTL-UPPAAL}), it does not support timed temporal operators such as $F^{[u,l]}$. 
This observation holds for most patterns of the catalog by \citet{AutiliGLPT15}.
Given the 18 pattern variants for qualitative and 19 variants for real-time properties, there exist 37 pattern variants. Each variant can be further configured by the scope, for which five options exist. This amounts to a total of 185 pattern variants.
Out of these 185 variants, only \textit{four} variants can be directly expressed in UPPAAL.
This drastically limited support of UPPAAL for the patterns is mainly caused by two issues. 
First, the four scopes apart from \textit{Globally} and the order patterns independently lead to formulae with nested temporal operators.
Second, the real-time property patterns require timed temporal operators.
Nested temporal operators and timed operators cannot be expressed with the query language of UPPAAL to formalize TCTL properties (\cf~Section~\ref{sec:uppaal-tctl}).

\subsection{Techniques to cope with the limited support of TCTL by UPPAAL}\label{sec:havelund}

\citet{HavelundLS99} have proposed conceptional and manual techniques to cope with the limited support of TCTL by UPPAAL (\cf~Section~\ref{sec:uppaal-tctl}). These techniques annotate the system model, define observer automata, and adjust the formulae for verification, which together encode those aspects that cannot be expressed by the query language of UPPAAL. Being conceptual, the techniques need to be concretized for the property at hand. 
In the following, we discuss these techniques, namely the \textit{flag}, \textit{debt}, and \textit{observer} techniques.

\subsubsection{The flag technique}

This technique adds a boolean \textit{flag} variable to the automata modeling the system, which is initialized to false and set to true if a designated state (in terms of a node in an automaton) is reached. 
The variable is then used in a formula to check whether the designated state has been reached while checking for the occurrence of another state. Thus, we can check for occurrences of paths from the designated to the other state, which would otherwise require a nested (T)CTL formula.

An example by \citet{HavelundLS99} is to consider an automaton with, amongst others, two states $a$ and $b$, and to verify that there is a path from $a$ to $b$. This property formalized in (T)CTL is $EF (a\ \wedge\ EF b)$, which cannot be expressed in UPPAAL due to the nesting of modal operators. Adding the flag variable \texttt{a\_reached} to the automaton that is assigned the value true when state $a$ has been reached (the assignment can be made on all incoming or outgoing transitions of state $a$), we can formulate the property as $EF(b\ \wedge\ a\_reached)$ and correspondingly in UPPAAL as $\texttt{E<>(b and a\_reached)}$. Thus, we check if there is a path where $a$ has been traversed before $b$ is reached. 

\subsubsection{The debt technique}

This technique extends the flag technique. It uses the metaphor of \textit{debt} to specify that if a certain state~\textit{a} is reached on the path to a target state~\textit{b}, then in-between a designated state~\textit{x} should be passed. Thus, we get into debt when reaching state \textit{a} and clear the debt when reaching state \textit{x}. The debt is represented by a boolean variable added to the automata. 

An example given by \citet{HavelundLS99} is to consider an automaton containing, amongst others, three states, $a$, $b$, and $x$ and to verify that every path from $a$ to $b$ must pass through $x$. Note, that the property is only concerned with paths that contain an \emph{a} before an occurrence of a \emph{b}. This property formalized in (T)CTL is $AG(a \rightarrow (\neg b\ W\ x))$---with $W$ being the \textit{weak until} operator---which cannot be expressed in UPPAAL due to unavailability of the \textit{W} operator in UPPAAL and the nesting of operators.
Adding the variable \texttt{debt} to the automaton, whose initial value is false, set to true when reaching state $a$ (\ie, we get into debt), and set again to false when reaching state $x$ (\ie, we clear the debt), we can formulate the property as $AG(b \rightarrow \neg debt)$ and correspondingly in UPPAAL as $\texttt{A[] b imply not debt}$. Thus, we check that if at any time state $b$ is reached, then there should be no debt because otherwise it would mean that state $a$ had been reached but not afterwards state $x$. 

\subsubsection{The observer technique}

This technique provides more flexibility than the other techniques as it does not limit the number of states whose reachability should be checked in particular orders. To achieve this, it adds communication actions to the automata modeling the system that synchronize with an added \textit{observer} automaton.\footnote{Since a transition of a timed automaton in UPPAAL can perform at most one communication action, adding such actions to the automaton typically requires adding transitions and corresponding states (nodes) to interact with the observer automaton.} Such actions notify the observer when designated states are reached. Thus, the observer monitors the system execution and particularly the order of designated states being reached while measuring time if needed.

An example by \citet{HavelundLS99} is to consider an automaton with, amongst others, two states, $a$ and $b$, and to verify that whenever state $a$ has been reached, then state $b$ must be reached from state $a$ within $t_u^b$ time units. This property formalized in TCTL is $AG(a \rightarrow AF^{[0,t_u^b]} (b))$, which cannot be expressed in UPPAAL due to the lack of timed operators such as $F^{[t_l,t_u]}$. Using the observer technique, we add two communication actions to the automaton. When reaching state $a$, a \texttt{begin!} signal is sent to the observer, and when reaching state $b$ an \texttt{end!} signal is sent. After receiving the \texttt{begin?} signal, the observer automaton resets its local clock to measure time and moves to another state to wait for the \texttt{end?} signal. If the \texttt{end?} signal arrives before $t_u^b$ time units, the observer moves to a ``\texttt{good}'' state, otherwise it moves to a ``\texttt{bad}'' state after $t_u^b$ time units have passed. Accordingly, the original property can now be formalized as a property of the observer as $AG\ \neg bad$ in TCTL and correspondingly as $\texttt{A[] not bad}$ in UPPAAL. Thus, we check that the observer will never reach state \texttt{bad}, in other terms, the signal \texttt{end?} will be always received within $t_u^b$ time units after receiving the signal \texttt{begin?}.

\section{A Property Specification Pattern Catalog for UPPAAL}\label{sec:Proposal}

In this section, we discuss our property specification pattern~(PSP) catalog for the verification of real-time systems with UPPAAL. 
The catalog comprises the patterns for qualitative and real-time properties collected by \citet{AutiliGLPT15}, and \textit{automatically} generates observer automata and related TCTL formulae from pattern-based specifications of properties. Using the catalog eases the specification of properties since practitioners specify properties in a pattern-based way without using a temporal logic or having to specify observer automata. 
Moreover, the catalog is seamlessly integrated with UPPAAL. The generated observer automata are UPPAAL-based timed automata with their related UPPAAL-compatible TCTL formulae. This enables the verification of the designated properties with UPPAAL.
Consequently, our comprehensive catalog bridges the gap between theoretical property specification patterns and practical model checking with existing tools, particularly the widely-used UPPAAL. 

\subsection{Overview}\label{sec:overview}

\begin{figure}[htb]
	\centering\includegraphics[width=1\linewidth]{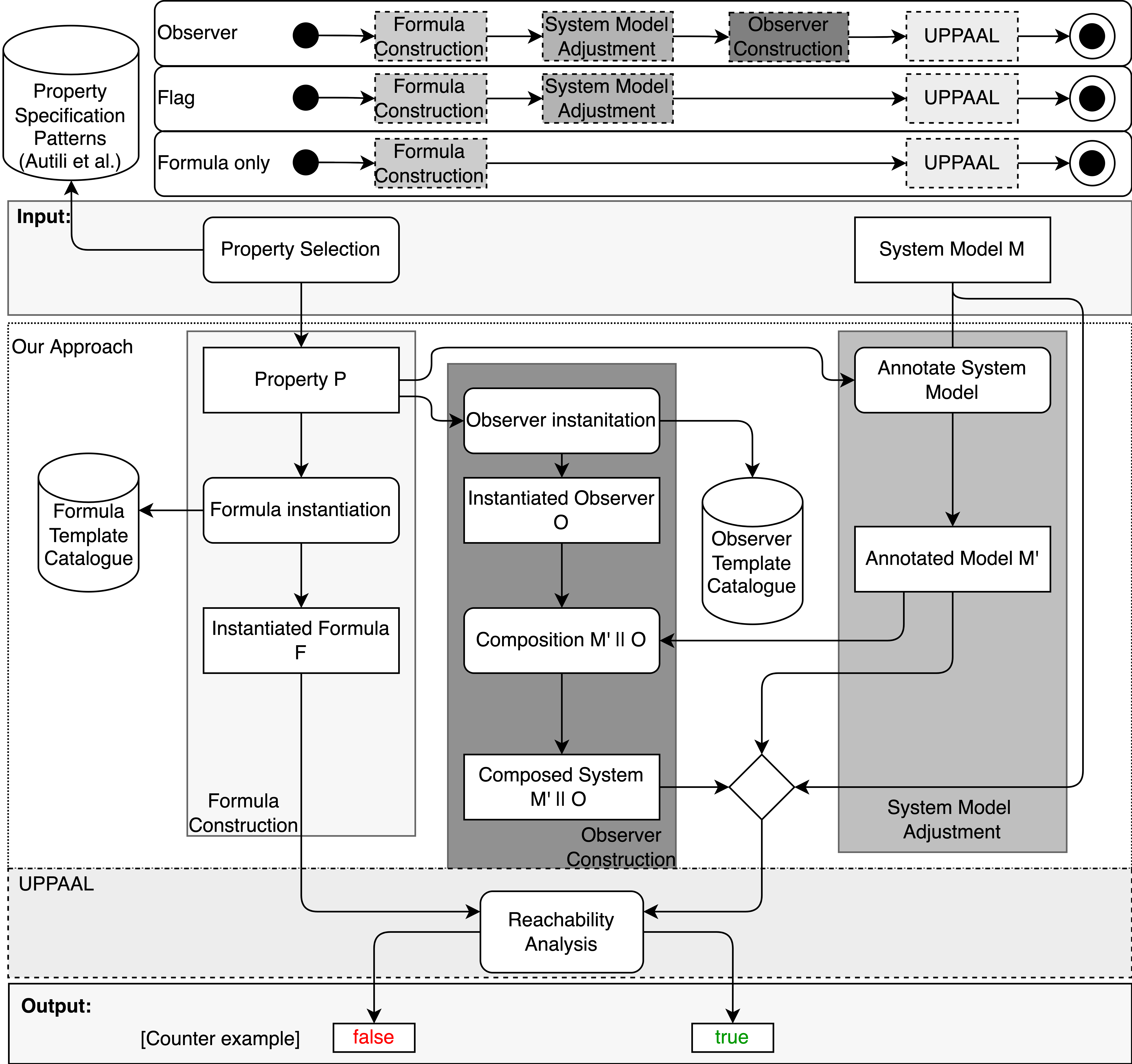}
	\caption{Overview of the end-to-end verification process of our pattern catalog.}
	\label{fig:psp_overview}
	\vspace{-1em}
\end{figure}

Figure~\ref{fig:psp_overview} shows an overview of the verification process of our catalog. Particularly, the catalog provides observer templates for the patterns, which are automatically instantiated to a given property defined in a pattern-based way. For each property, the generated observer and related formula to be verified are automatically composed with the user-defined UPPAAL model of the system (timed automata specification models describing the target-system behavior) in order to perform the verification in UPPAAL. Thus, our catalog with its integration to UPPAAL enables a verification process from the pattern-based specification to the model checking in UPPAAL.

Following Figure~\ref{fig:psp_overview}, the verification process of our pattern catalog is comprised into three major steps: (i) \textit{Formula Construction}, (ii) \textit{System Model Adjustment}, and (iii) \textit{Observer Construction}. In each step, we apply a particular technique to be able to verify properties based on the patterns of the PSP catalog for UPPAAL.

In the \emph{Formula Construction} step, we translate the PSP formula into the UPPAAL property specification language by instantiating the corresponding template in the \textit{Formula Template Catalog} (Figure~\ref{fig:psp_overview}). We further explain how this process is carried out in our approach and present our Formula Template Catalog in Section~\ref{sec:formula}. In the \textit{System Model Adjustment} step, we apply both the flag and the debt techniques~\cite{HavelundLS99} to the original system model M after the adequate formula is instantiated from the Formula Template Catalog (Figure~\ref{fig:psp_overview}). Given that the debt technique is an extension of the flag technique, we will refer to the application of the flag technique in the System Model Adjustment step for simplicity. We further delve into the System Model Adjustment step in Section~\ref{sec:flag}. Finally, in the \emph{Observer Construction} step, the observer template is instantiated. Whenever applicable, there is a corresponding observer in the \textit{Observer Template Catalog} for each property pattern of the PSP catalog of Autili et al.~\cite{AutiliGLPT15}. We further discuss the Observer Construction step in Section~\ref{sec:observer}.

\begin{table*}[t!]
	\centering
	\caption{Schematic view of the process applied in our approach to the property classification according to the catalog by \citet{AutiliGLPT15}.}  
	\label{tab:UPPAALPSP}
	\vspace{-1em}
	\footnotesize
	\begin{tabular}{crlccc}
		\toprule
		\textbf{Type} & \textbf{\#} & \textbf{Pattern Name} & \textbf{Scope Name} & \textbf{Qualitative} & \textbf{Real-time} \\ \midrule
		\multirow{13}{*}{\rotatebox{90}{Occurence}} 
		&  & \multirow{3}{*}{Absence} & Globally & Formula & Formula\\
		& 1 & & After & Flag & Observer\\
		& & & All the others & Observer & Observer \\\cline{2-6}
		&  & \multirow{3}{*}{Universality} & Globally & Formula & Formula\\
		& 2 & & After & Flag & Observer\\
		& & & All the others & Observer & Observer \\\cline{2-6}
		&  & \multirow{3}{*}{Existence} & Globally & Formula & Formula\\
		& 3 & & After & Flag & Observer\\
		& & & All the others & Observer & Observer \\\cline{2-6}
		& 4 & Bounded Existence & All scopes & Observer & N/A \\\cline{2-6}
		& 5 & Minimum Duration & All scopes & N/A & Observer\\\cline{2-6}
		& 6 & Maximum Duration & All scopes & N/A & Observer\\\cline{2-6}
		& 7 & Recurrence & All scopes & Observer & Observer\\\midrule
		\multirow{15}{*}{\rotatebox{90}{Order}}
		& 8 & Precedence & All scopes & Observer & N/A \\\cline{2-6}
		& 9 & Precedence Chain$_{N-1}$ & All scopes &Observer & N/A \\\cline{2-6}
		& 10 & Precedence Chain$_{1-N}$ & All scopes &Observer & N/A \\\cline{2-6}
		& 11 & Constrained Precedence Chain$_{N-1}$& All scopes &Observer & N/A \\\cline{2-6}
		& 12 & Constrained Precedence Chain$_{1-N}$ & All scopes &Observer & N/A \\\cline{2-6}
		& \multirow{2}{*}{13} & \multirow{2}{*}{Response} & Globally & Formula & Observer\\
		& & & All the others & Observer & Observer\\\cline{2-6}
		& 14 & Response Chain$_{N-1}$ & All scopes &Observer & Observer \\\cline{2-6}
		& 15 & Response Chain$_{1-N}$ & All scopes &Observer & Observer \\\cline{2-6}
		& 16 & Constrained Response & All scopes &Observer & Observer \\\cline{2-6}
		& 17 & Constrained Response Chain$_{N-1}$ & All scopes &Observer & Observer \\\cline{2-6}
		& 18 & Constrained Response Chain$_{1-N}$ & All scopes &Observer & Observer \\\cline{2-6}
		&\multirow{2}{*}{19} & \multirow{2}{*}{Response Invariance} & Globally & Flag & Observer\\
		& & & All the others & Observer & Observer\\\cline{2-6}
		& 20 & Until & All scopes &Observer & Observer \\ \bottomrule
	\end{tabular}
	
\end{table*}

Indeed the major process of our approach is to have all three aforementioned steps towards constructing timed observer automata, namely the \emph{Observer process}. However, some property patterns may be verified in UPPAAL with the Formula Construction step either alone, namely the \emph{Formula-Only process}, or followed by the System Model Adjustment step, namely the \emph{Flag process}, without requiring to go further into the Observer Construction step. These three sub-processes are depicted on top of Figure~\ref{fig:psp_overview}. The decision on which process to apply depends on the feasibility of expressing the property specification pattern directly in UPPAAL's TCTL formula. In Table~\ref{tab:UPPAALPSP} we summarize how each of those processes maps to the corresponding property pattern and scope. We should note that out of the 185 combinations of patterns (18 qualitative and 19 real-time patterns) and scopes (five scopes) we investigated\footnote{A real-time variant of the \textit{Bounded Existence} pattern and qualitative variants of the \textit{Minimum} and \textit{Maximum Duration} patterns do not exist in general (Table~\ref{tab:UPPAALPSP}).}, only four apply the Formula-Only process. The Flag process was applied to four combinations of patterns and scopes for qualitative, i.e., untimed, properties. All of the remaining combinations applied the Observer process. We also highlight the fact that the real-time variants of the \textit{Precedence}, \textit{Precedence Chain}, and \textit{Constrained Precedence Chain} patterns are not applicable in UPPAAL (Table~\ref{tab:UPPAALPSP}), which results in 160 combinations of patterns and scopes that we realized for our catalog. Using the Structured English Grammar~\cite{AutiliGLPT15}, the Precedence pattern is described as:  

\begin{quote}
\textbf{Scope}, \textbf{if} \textit{\{P\}} \textbf{[holds]}, \textbf{then it must have been the case that} \textit{\{S\}} \textbf{[has occurred]} \textbf{[Interval(0)]}  \textbf{before} \textit{\{P\}} \textbf{holds}. 
\end{quote}

Simply stated, the reason behind is that any instance of \textit{S} would have to be taken into consideration when evaluating an instance of \textit{P}. UPPAAL, however, cannot handle arbitrary many variables, states, or arrays with arbitrary length. We should note that if the maximum number $n$ of occurrences of S for any given point in time is known \emph{a priori}, then $n$ observers can be used in parallel to express the property.

\subsection{The Formula Construction Step}\label{sec:formula}

In this first part of the process, we fetch the provided property and instantiate the corresponding UPPAAL formula template from Autili et al.~\cite{AutiliGLPT15}\footnote{As previously mentioned, in this work we focus only on qualitative and real-time properties as UPPAAL does not support probabilistic quantifiers.}. Most of our templates are expressed as safety properties. This means that during the verification step, it is sufficient for UPPAAL to show that an unwanted state, mostly identified as an \verb+ERROR+ location, is not reachable.
Some properties, however, are expressed as liveness properties, for instance, the Response pattern. In this case, the observer is used to check whether the specified response happens eventually.
A violation of the specification is detected as a counterexample when the observer can reach such an \verb|ERROR| location. We will further delve into the observer automata for the patterns in Section~\ref{sec:observer}.

Our UPPAAL formula template catalog includes 18 qualitative and 14 real-time patterns. Additionally, each pattern must use one of five scopes. Therefore, our formula catalog amounts to 160 formula templates\footnote{For the sake of readability of all those 160 formula templates, we refer the reader to our Github repository at: \url{https://github.com/hub-se/PSP-UPPAAL}}. 

Out of those formula templates there are only four that can be directly verified without the need of observer automata or any system model adjustments. We call this special case the \emph{Formula-Only} process as only the formula is needed for verification. It can be applied for the \textit{Globally} scope paired with the \textit{Absence}, \textit{Universality}, and \textit{Existence} patterns, both in qualitative and real-time variants. Furthermore, the qualitative version of the \textit{Globally} scope paired with the \textit{Response} pattern can be expressed with a formula using UPPAAL's \emph{leads-to} (\verb|-->|) operator. The reason behind this is that the formulae of such patterns are the only ones that are not nested (considering the catalog by \citet{AutiliGLPT15}), except for the Response pattern. For real-time properties, we automatically add a global clock $gc$ to the system. Afterwards, UPPAAL's reachability analysis can be performed.

\subsection{The System Model Adjustment Step}\label{sec:flag}

The next step in the process is called the \emph{System Model Adjustment}, where we apply both the flag and the debt techniques (henceforth referred only as flag, for short) proposed by Havelund et al.~\cite{HavelundLS99}. To enable the interaction between the system model and an observer, the system model has to be adjusted. For this purpose, we apply the System Model Adjustment step where not only we make traceable the states that ought to be verified as part of the property but also make traceable the paths in which the state, or set of states, of interest is passed through. In Figure~\ref{fig:psp_overview}, the adjusted model is represented by $M'$. Once the pattern type and scope have been identified from the user's input, there are only two steps required: one to fetch the location in the system model M that represents the state of interest and the other to apply the flag technique to that location.

\begin{figure*}[t!]
	\centering
	\begin{subfigure}[t]{0.5\textwidth}
		\centering
		\includegraphics[scale=.5]{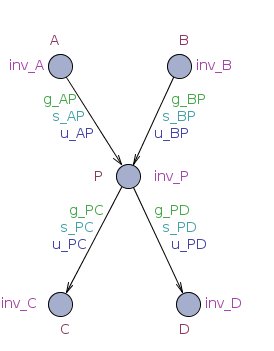}
		\caption{}
		\label{fig:psp_flag_Prior}
	\end{subfigure}%
	~ 
	\begin{subfigure}[t]{0.5\textwidth}
		\centering
		\includegraphics[scale=.5]{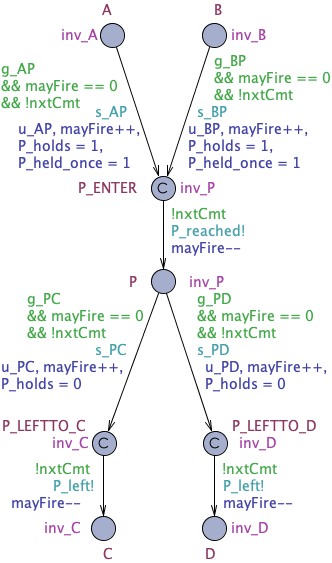}
		\caption{}
		\label{fig:psp_flag_After}
	\end{subfigure}
	\caption{Illustration of the System Model Adjustment step. (a) The model prior to applying the \texttt{applyFlag()} algorithm. (b) The outcome of the algorithm with the extended location \texttt{P} following the flag and debt techniques (Section~\ref{sec:Background}). Moreover, the model is also adjusted to interact with an observer, thus supporting the observer technique (Section~\ref{sec:Background}).}
	\label{fig:psp_flag}
\end{figure*}

The core of our System Model Adjustment step is presented in Algorithm~\ref{func:applyDebt}. We explain this algorithm as follows in tandem with Figure~\ref{fig:psp_flag} to help the reader depict the intricate System Model Adjustment step. In Figure~\ref{fig:psp_flag}, we illustrate the effect of applying the \texttt{applyFlag()} method of Algorithm~\ref{func:applyDebt} into a certain state \texttt{P}\footnote{Note that \texttt{P} must be a traceable location of the original model M. Otherwise, one must refactor M by manually adding \texttt{P} prior to the System Model Adjustment step.}. The original model excerpt regarding \texttt{P} is presented in Figure~\ref{fig:psp_flag_Prior}, while the outcome of the \texttt{applyFlag()} into that excerpt is illustrated in Figure~\ref{fig:psp_flag_After}.

\begin{algorithm}[!htb]
	\footnotesize
	\caption{applyFlag()}
	\label{func:applyDebt} 
	\begin{algorithmic}[1]
		\Require System\_model M, Location P \label{algline:input}
		\Ensure Adjusted\_model M' 
		
		\ForAll{Transition T \textnormal{\textbf{in}} M}
		\State T.addGuard(mayFire == 0);
		\EndFor
		\State addLocation(P\_ENTER);
		\State P\_ENTER.setCommitted();
		\State P\_ENTER.setInvariant(P.invariant);
		\State addTransition(P\_ENTER, P);
		\State P\_ENTER.getTransition(P).setSynchronisation(P\_reached!);
		\State P\_ENTER.getTransition(P).setUpdate(P\_holds = 1, P\_held\_once = 1);
		
		\ForAll{Location S \textnormal{\textbf{in}} P.predecessors}
		\State S.getTransition(P).adjustTransitionTo(P\_ENTER);
		\EndFor
		
		\ForAll{Location T \textnormal{\textbf{in}} P.successors}
		\State addLocation(P\_LEFTTO\_T);
		\State P\_LEFTTO\_T.setInvariant(T.invariant);
		\State P.getTransition(T).adjustTransitionTo(P\_LEFTTO\_T);
		\State addTransition(P\_LEFTTO\_T, T);
		\State P\_LEFTTO\_T.getTransition(T).setSynchronisation(P\_left!);
		\State P\_LEFTTO\_T.getTransition(T).setUpdate(P\_holds = 0);
		\State P\_LEFTTO\_T.setCommitted();
		\EndFor
		
		\ForAll{Transition T \textnormal{\textbf{in}} M}
		\State T.addGuard(!nxtCmt);
		\EndFor
		
		\State \Return M'
	\end{algorithmic}
\end{algorithm}

To be able to trace incoming and outgoing paths of \texttt{P}, we add synchronizations \texttt{P\_reached} and \texttt{P\_left}. Since \texttt{P} can be entered or left as a result of a synchronization (e.g., \texttt{s\_AP} or \texttt{s\_PC}) and UPPAAL only allows a single synchronization per transition, we add locations as pseudo-states before and after \texttt{P}. We use UPPAAL's committed locations to ensure that time cannot pass while the system is in a state containing such a location and that the system model’s execution is not altered by the addition of these locations. If one of the system’s automata is in a committed location, a transition exiting the committed location must be taken immediately upon entering it. 
The committed location of the incoming transitions to \texttt{P} is named \texttt{P\_ENTER}, cmp. lines 4-7, Algorithm~\ref{func:applyDebt}.
The newly added location \texttt{P\_ENTER} is connected to \texttt{P} by a transition which synchronizes with the observer via the \texttt{P\_reached} channel, line 8 of Algorithm~\ref{func:applyDebt}. We use UPPAAL's ability of broadcast channels since these are non-blocking, i.e., an observer does not have to synchronize in order for the system model to use the transition. Furthermore, multiple observers could synchronize with the one sender; in fact every observer that \emph{can} synchronize, \emph{must} do so.

Using synchronizations when a state of interest is entered enforces progress 
in the observer. Sometimes, however, the observer cannot synchronize immediately but instead needs to know later on if a certain state had been reached before. Therefore, we use the added locations representing pseudo-states to update boolean flags with these information: \texttt{P\_holds} and \texttt{P\_held\_once}. Essentially, \texttt{P\_holds} when the system is in state \texttt{P}, and \texttt{P\_held\_once} holds whenever the state \texttt{P} was reached at least once.  As such, both are set to $1$ when \texttt{P} is entered, cmp.\@ line 9, Algorithm~\ref{func:applyDebt}. This update does not block or alter the system's behavior, as long as no such variable is used by the system originally. The first variable (\texttt{P\_holds}) follows the principle of a \textit{debt} variable (Section~\ref{sec:Background}) and is a paramount variable that comprises a great number of our observer templates to check whether the paths passing through \texttt{P} have been traversed. Additionally, the \texttt{P\_held\_once} variable applies the principle of a \textit{flag} variable (Section~\ref{sec:Background}). This flag is mostly used by our formula templates that compose the \textit{After} scope in tandem with the \textit{Occurrence} patterns to trace those paths that have reached \texttt{P} once. All transitions entering \texttt{P} will then be redirected to \texttt{P\_ENTER}, as can be seen in lines 10-12 of Algorithm~\ref{func:applyDebt}. Between \texttt{P} and each of its outgoing state such as \texttt{C}, we need to add another pseudo-state represented by the location \texttt{P\_LEFTTO\_C} and copy its corresponding invariant (e.g., \texttt{inv\_C}). Otherwise, the transition into the pseudo-state may be enabled while the transition out of the \texttt{P\_LEFTTO\_C} pseudo-state may be disabled, which could render a deadlock. When leaving the pseudo-state, we can again provide a broadcast synchronization for the observer, i.e., \texttt{P\_left}, and an update, i.e., \texttt{P\_holds = 0} (cmp. also lines 13-21 of Algorithm~\ref{func:applyDebt}).

We should note that, to ensure that no other system transition may be enabled while $M'$ is in a pseudo-state, we add a semaphore variable \texttt{mayFire}, cmp. lines 1-3 of Algorithm~\ref{func:applyDebt}. Every transition in the model is updated with a guard condition to disable it while the system is in a pseudo-state. We decided to use an integer instead of a boolean value for the semaphore because the system may enter multiple locations all representing pseudo-states in one step. Using an integer enables us to increment it for each pseudo-state that is entered and decrement it for each pseudo-state that is left. Additionally, to allow the observer to prioritize its own transitions, we add another semaphore: \texttt{nxtCmt}, cmp. lines 22-24 of Algorithm~\ref{func:applyDebt}. For this purpose, the observer sets \texttt{nxtCmt = true}, which disables every system transition extended with the guard condition \texttt{!nxtCmt}. 
Furthermore, for the timed patterns we add a global clock variable to the system that we use in our formulae and in their corresponding observers. 

As previously mentioned in Section~\ref{sec:overview}, some patterns may go into the Formula Construction step, followed by the System Model Adjustment step. We call this the \emph{Flag process} in our approach, where the application of the flag technique suffices and no observer construction is needed. This is the case for the patterns of \textit{Absence}, \textit{Universality,} and \textit{Existence} combined with the scope \textit{After}.

\subsection{The Observer Construction Step}\label{sec:observer}

The third and final step in our approach is the \textit{Observer} process. Intuitively, observers (represented as $O$ in Figure~\ref{fig:psp_overview}) run in parallel with the model under verification, in our case, the adjusted model M' as an outcome of the previously explained System Model Adjustment step. 

Following from Figure~\ref{fig:psp_overview}, the observer automaton that will compose with the adjusted model M' is an instantiation from our Observer Template Catalog based on the property pattern of interest and its corresponding scope. Therefore, each combination of pattern and scope holds a unique observer template in our catalog, except for those that follow the Formula-Only or Flag processes, as previously explained. 

\begin{figure*}[t]
	\centering
	\includegraphics[scale=.5]{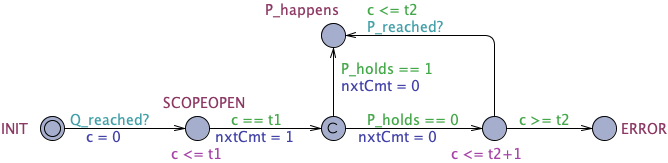}
	\caption{Observer of P state for the Timed Existence pattern with the scope \textit{After Q}.}
	\label{fig:Existence_Observer}
\end{figure*}

The principle behind the design of each observer template follows initially the observer technique by Havelund et al.~\cite{HavelundLS99}, where the synchronizations \texttt{P\_reached} and \texttt{P\_left} follow their principle of the \texttt{begin} and \texttt{end} channels, respectively. In addition, considering (1) the property pattern intent, (2) the scope, and (3) the pattern category (i.e., qualitative or real-time), we designed each observer template to represent those three elements that comprise the synthesis of the observer from the template. 

All in all, the instantiation of the appropriate observer template for its corresponding property can be briefly described in two steps: (1) fetching the corresponding observer template from the catalog and (2) replacing the template's variables with actual system states and values for time bounds and integers. 

As described earlier, our properties consist of safety and liveness properties. For safety properties, the observed automata must reach a safe location in the reachability analysis process. A violation of a specification is detected when an unwanted location can be reached in the observer, identified as an \texttt{ERROR} location. A violation of the specification is detected as a counterexample when the observer can reach such a location. As for liveness properties (e.g., ``every occurrence of P is followed by an occurrence of Q''), we use acceptance conditions for this purpose: some locations in the observer are marked as accepting locations. Reasoning about infinite runs is also necessary. For infinite runs, a counterexample is detected if there is any run that never satisfies the accepting location.

To illustrate that principle, suppose we want an observer for the model illustrated in Figure~\ref{fig:psp_flag_After} for the timed \textit{Existence} property pattern with the scope \textit{After}, described with the Structured English Grammar~\cite{AutiliGLPT15} as follows:

\begin{quote}
 \centering
 \textbf{After} \emph{Q}, \emph{P} \textbf{holds eventually} \emph{Time(P)}
\end{quote}

This property pattern aims at describing a portion of a system's execution during time interval t1 $\leq Time(P) \leq$ t2  that contains an instance of certain state P \emph{after} another state Q happened. 
The observer template that models this expected pattern behavior is shown in Figure~\ref{fig:Existence_Observer}, where the observer will only start to observe once the event \texttt{Q\_reached} has been triggered by the adjusted model M' of interest. At that moment, the clock \texttt{c} is reset to start counting elapsed time. The final states of the observer are either \texttt{P\_happens} or \texttt{ERROR} for the wanted and unwanted behaviors, respectively. 
\texttt{P\_happens} will be reached in two cases: Either when $P$ already holds when opening the time interval (timed scope), or when $P$ is reached after the interval has opened, but before it closes. In the case that the time bound has been violated, the \texttt{ERROR} state will be reached showing that the property has been violated. 
The verification with such an observer requires simply the following liveness property: \texttt{SCOPEOPEN --> P\_happens}, expressing that whenever \texttt{SCOPEOPEN} holds, it is always the case that \texttt{P\_happens} will eventually hold within the time bound [t1,t2]. Finally, once the actual system property with its instantiated values for \texttt{P} and \texttt{Q} are known, the observer model will be then automatically instantiated replacing all the variables and channels related to \texttt{P} and \texttt{Q} accordingly.

\subsection{Illustrating the Approach}

To exemplify our application on the BSN example, let us follow property BSN-P04 from Table \ref{tab:bsn-properties}. It specifies that at most one scheduler cycle ends before an emergency is detected if the patient's health status is at high risk. We mapped this property to the \textit{Bounded Existence} pattern, investigating if at most one scheduler cycle elapses within the two events ``health status being high" and "detecting an emergency". We use these two events to set the \textit{Between} scope:
\begin{quote}
\textbf{Between} \emph{the bodyhub status is high} \textbf{and} \emph{emergency is detected}, \emph{the scheduler cycle is finished} \textbf{holds at most} \emph{1} \textbf{times.}
\end{quote}

The first step in our application is to load the formula corresponding to the pattern and scope. This formula is instantiated by replacing placeholders with the actual system states and, in this case, the variable $n$ with the concrete value $1$. 

\begin{figure*}[ht!]
	\centering
	\begin{subfigure}[t]{\textwidth}
		\centering
		\includegraphics[scale=.5]{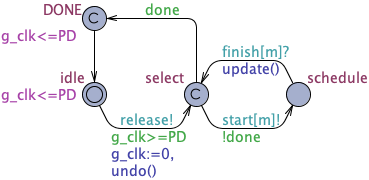}
		\caption{}
		\label{fig:BSNP04:schedulerBefore}
	\end{subfigure}%
	
	\begin{subfigure}[t]{0.55\textwidth}
		\centering
		\includegraphics[scale=.5]{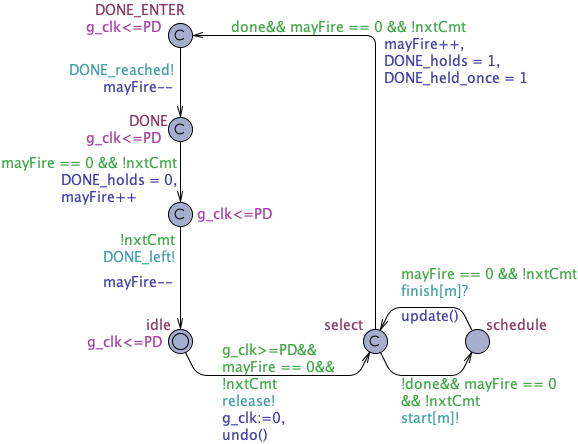}
		\caption{}
		\label{fig:BSNP04:schedulerAfter}
	\end{subfigure}
	\begin{subfigure}[t]{0.4\textwidth}
		\centering
		\includegraphics[scale=.5]{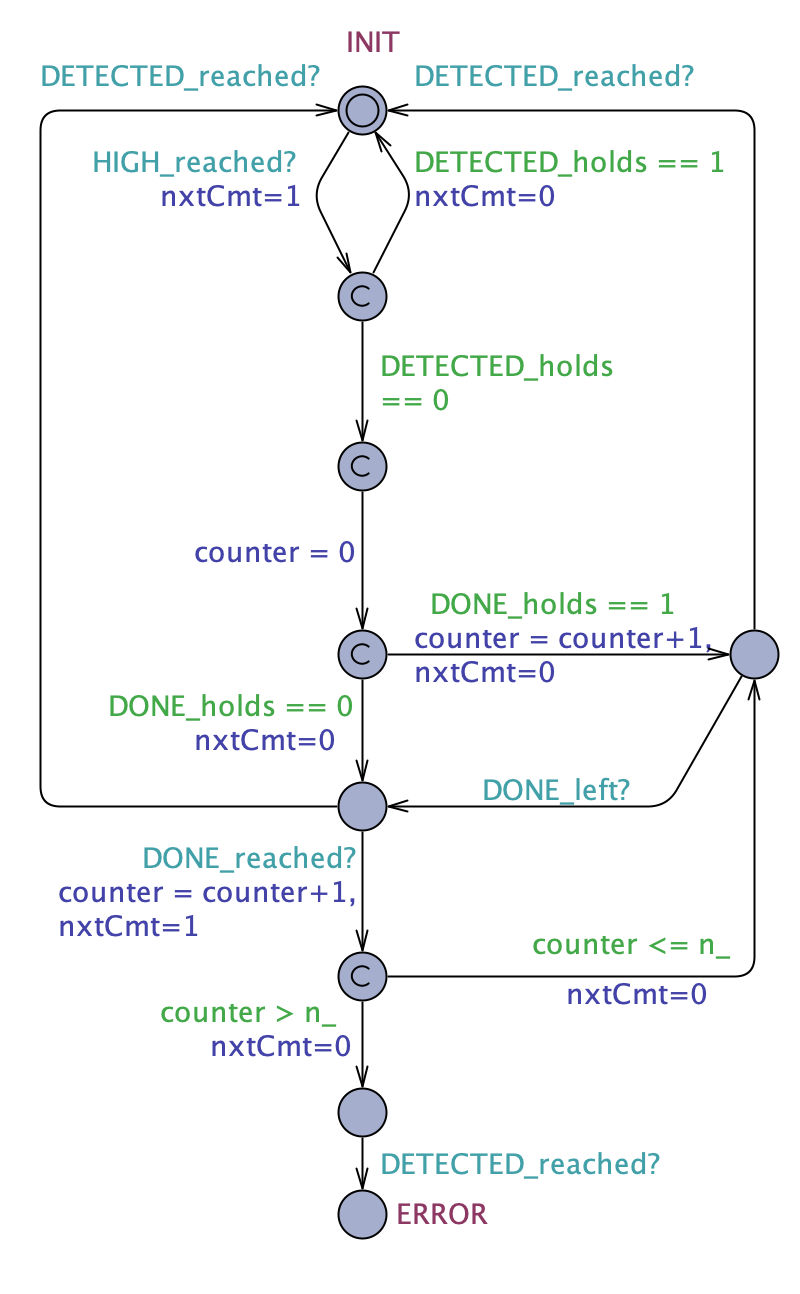}
		\caption{}
		\label{fig:BSNP04:observer}
	\end{subfigure}
	\caption{The BSN UPPAAL automata for property BSN-P04 and its observer: the automaton prior to (Figure~\ref{fig:BSNP04:schedulerBefore}) and after (Figure~\ref{fig:BSNP04:schedulerAfter}) the Observer for property P04 (Figure~\ref{fig:BSNP04:observer}) was generated through the Observer process of our approach.}
	\label{fig:BSNP04}
	\vspace{-1em}
\end{figure*}

Afterwards, the system model is loaded and adjusted. The locations in the model that represent the system states used in the property are annotated with synchronizations and flags, as described previously in the context of Figure \ref{fig:psp_flag} and Algorithm \ref{func:applyDebt}.
Figures~\ref{fig:BSNP04:schedulerBefore} and~\ref{fig:BSNP04:schedulerAfter} display the system automaton of the scheduler before and after the annotations were added. 

Furthermore, the template for the observer automaton is loaded and instantiated, similarly to the formula. Placeholders are replaced with corresponding synchronizations and values for variables. The resulting observer automaton is depicted in Figure~\ref{fig:BSNP04:observer}.

In Figure \ref{fig:BSNP04:observer}, the property of interest is broken down with an observer. The formula for the Bounded Existence pattern with the Between scope directly transcribed from the PSP catalog by \citet{AutiliGLPT15} is as follows:
\begin{quote}
   $AG (Q\rightarrow \neg E[\neg R \mathcal{U} (\neg P \wedge \neg R \wedge EX( P \wedge E [\neg R \mathcal{U} (\neg P \wedge \neg R \wedge EX(P \wedge E[\neg R \mathcal{U} (\neg P \wedge \neg R \wedge EX (P \wedge\neg R \wedge EF(R)))]))]))])$
\end{quote}

Given the fact that the CTL formula obtained from the PSP catalog contains many nested temporal operators, it cannot be directly checked in UPPAAL as it is. Our observer, however, encodes information about the different parts of the formula into states. Consider, for example, the first implication: It expresses that $Q$ opens the scope. In the example, \texttt{HIGH} is a concrete state of the system that substitutes $Q$. It encodes the information that the patient's health status was detected as \emph{high}. In the observer's initial state, the transition can only be enabled once the state \texttt{HIGH} is reached in the system resulting in a synchronization between the adjusted system model and the observer over the channel \texttt{HIGH\_reached}. Then, if \texttt{DETECTED\_holds} is evaluated to $1$, the scope is considered to be closed, because \texttt{DETECTED\_holds} depicts the variable $R$, representing that an emergency has been detected. Only if \texttt{DETECTED\_holds} is evaluated to $0$, meaning no emergency has been detected yet, the scope is finally open and a counter is being reset. Every occurrence of \texttt{DONE}, meaning a scheduler cycle has finished, increments the counter. If the emergency is detected, the scope will be closed again. If, however, the scheduler cycle ends repeatedly, i.e., more than once, the observer reaches the state~\texttt{ERROR}. 

\section{Evaluation}\label{sec:Evaluation}

We evaluate our catalog together with its generator for UPPAAL-based observers and formulae on three real-time systems from literature~\cite{LindahlPY01,HavelundSLL97, RodriguesCR0P18}, two of which have been applied in an industrial setting. These systems have been originally verified with UPPAAL using manually created observer automata. Applying our catalog to these systems, we demonstrate that our catalog can express the properties of interest and generate corresponding observer automata and formulae. Using these generated observer automata and formulae, we were able to reproduce the verification results reported in literature for these systems, which provides evidence for the validity of our catalog.
The major goal of this experiment is to assess the ability of our framework to empirically demonstrate that the generation of our observers is both feasible and preserves the behaviors of interest in an existing UPPAAL model. By doing so, we are able to leverage not only the ability to automatically generate observers from pattern-based properties but also to benefit from an observer-based verification process as well as the property specification pattern approach.

To conduct our experimentation, we hypothesize that our automated approach yields the same behavior by satisfying the same properties as those obtained by manually formulating UPPAAL queries. To assess our hypothesis, we perform an experimentation with three UPPAAL systems reported in the literature: the Body Sensor Network~\cite{RodriguesCR0P18}, the Audio/Video Protocol~\cite{HavelundSLL97} and the Gear Controller~\cite{LindahlPY01}. In particular, those systems were chosen as we could replicate their UPPAAL models as provided in their respective work. We follow a technology-oriented experiment~\cite{ExpSEBuch:2012} and the results and the UPPAAL models presented in this section are publically available\footnote{\url{https://github.com/hub-se/PSP-UPPAAL/wiki/Case-Studies}}. 

\subsection{Experimental Setup}\label{sec:setup}

For the evaluation we used UPPAAL in version 4.1.20 beta 3. To build the generator for observers, we used Apache Maven 3.6.3. The tool is publicly available in the project's repository\footnote{\url{https://github.com/hub-se/PSP-UPPAAL/}}. All dependencies are loaded automatically during the build with Maven to ensure reproducibility. The generator is implemented in Java version 1.8.0\_301. 

The experiments were conducted on a 3,1 GHz Quad-Core Intel Core i5 and 16 GB memory. For each experiment, we recorded the time it took our tool to analyze and modify the system model and the number of locations and transitions that were added. An extensive report of our results can be found in our repository, where we list the experiments' data and the number of added locations and transitions.

Note that our approach only handles requirements specifying information of locations in the UPPAAL model but not of variables. This is due to the fact that we need to enforce progress in the observer if the intended behavior manifests in the model. We do this in UPPAAL with synchronizations of locations (i.e., UPPAAL states) which is not feasible via variables. Therefore, we had to manually refactor the systems when variables were specified in the requirements so that the property specification could be expressed by means of locations in the UPPAAL model instead. Information about the changes we conducted can be also found in the repository. Furthermore, note that, when applicable, the timed versions are indicated explicitly in each system property.

\subsection{Experiments with the UPPAAL Systems}

In this section, we present the informal description of properties of the selected UPPAAL systems, followed by their description using the structured English grammar by Autili et al.~\cite{AutiliGLPT15}. Then we describe which process of our approach we followed to verify each property based on its according pattern. 

\subsubsection{Body Sensor Network}

\begin{table*}[!t]
	\footnotesize
	\caption{Properties of the BSN.}
	\label{tab:BSNproperties}
	\centering
	\begin{tabular}{p{0.08\textwidth}p{0.52\textwidth}p{0.26\textwidth}}
		\toprule
		\textbf{ID} & \textbf{Structured English Description} & \textbf{Property Pattern and Scope}\\ \toprule
		\textbf{BSN-P02} &  Between \emph{Scheduler.Done} and \emph{Scheduler.Idle, Bodyhub.Idle} eventually holds. & Existence Between \\ \midrule
		\textbf{BSN-P03} &  Between \emph{Scheduler.Done} and \emph{Scheduler.Idle, Sensornode.Idle} eventually holds. & Existence Between\\ \midrule
		\textbf{BSN-P04} &  Between \emph{Sensornode.HIGH} and \emph{Bodyhub.DETECTED}, \emph{Scheduler.DONE} holds at most one time. & Bounded Existence Between \\ \midrule
		\textbf{BSN-P05} &  Globally, if \emph{Sensornode.HIGH} has occurred, then in response \emph{Bodyhub.DETECTED} holds within \emph{250ms}. & Time-constrained Response Globally \\ \midrule
		\textbf{BSN-P06} &  Globally, if \emph{Sensornode.Request} has occurred, then in response \emph{Sensornode.Reply} eventually holds. & Response Globally\\ \midrule
		\textbf{BSN-P07} &  Globally, if \emph{Sensornode.Request} has occurred, then in response \emph{Sensornode.Reply} eventually holds. & Response Globally\\ \midrule
		\textbf{BSN-P08} &  Globally, if \emph{Sensornode.Request} has occurred, then in response \emph{Sensornode.Reply} eventually holds. & Response Globally\\ \midrule
		\textbf{BSN-P09} &  Globally, if \emph{Sensornode.COLLECTED} has occurred, then in response \emph{Bodyhub.PROCESSED} eventually holds. & Response Globally\\ \midrule
		\textbf{BSN-P10} &  Globally, if \emph{Sensornode.COLLECTED} has occurred, then in response \emph{Bodyhub.PERSISTED} eventually holds. & Response Globally\\ \midrule
		\textbf{BSN-P11} &  Globally, if \emph{Sensornode.SENT} has occurred, then in response \emph{Bodyhub.PR\_correctly} eventually holds. & Response Globally\\ \midrule
		\textbf{BSN-P12} &  Globally, if \emph{Bodyhub.PROCESSED} has occurred, then in response \emph{Bodyhub.DETECTED} eventually holds. & Response Globally\\ \bottomrule
	\end{tabular}
\end{table*}

The Body Sensor Network (BSN) has been previously explained in Section~\ref{sec:BSN}. The models as well as a list of requirements for the BSN are further described in \cite{RodriguesCR0P18}. Disregarding requirement \emph{BSN-P01}, which is only the check for deadlock-freedom, we verified all the remaining eleven requirements listed in Table~\ref{tab:bsn-properties}. 
The corresponding descriptions of those requirements in structured English as well as their property patterns and scopes are presented in Table \ref{tab:BSNproperties}.
The BSN system model consists of 27 locations and 40 transitions.

From the properties described in Table~\ref{tab:BSNproperties}, one can notice that both Occurrence and Order patterns are applied. While BSN-P02 and BSN-P03 are Existence properties with a Between scope, property BSN-P04 is a Bounded Existence with a Between scope. As such, those three properties follow our Observer process. BSN-P05 also follows the Observer process as it uses the time-constrained Response Globally pattern. Properties BSN-P06 to BSN-P12 are all Untimed Response properties with a global scope (Response Globally for short). Therefore, they go through the Formula-Only process.

\subsubsection{Audio/Video-Protocol}

Havelund et al.~\cite{HavelundSLL97} used UPPAAL to locate errors in an audio/video-protocol used in industry by Bang \& Olufsen. This protocol features a set of senders that send frames over a single bus. Naturally, collisions may happen here. The industry partner complained that there were faults in the system that they were not able to locate. Using the UPPAAL model checker to help them, \citet{HavelundSLL97} modeled the system as timed automata. The model consists of two senders, each modeled with four automata, and another automaton for the bus. To locate the faults, the authors hand-built an observer automaton that observes the system. Seventy-one locations and 133 transitions are used in the model, including the original observer. Through our approach, we were able to discard their hand-built observer automaton and instead, use the observers we generated automatically from our approach based on the corresponding property pattern. Verifying the adapted models located the same faults that were found and discussed by \citet{HavelundSLL97}. In Section~\ref{sec:results} we report the outcome of our experiments on the Audio/Video-Protocol in detail.

We describe in Table~\ref{tab:havelundproperties} the property patterns applied to the Audio/Video-Protocol system. It can be noticed that we only needed to apply the Response pattern with the Between scope. As such, AV-P01(a), AV-P01(b), and AV-P02 follow our Observer process. Note, that while the properties may appear similar, we adapted the underlying system model separately for each property.

\begin{table*}[!h]
    \footnotesize
	\caption{Properties of the Audio/Video Protocol.}
	\label{tab:havelundproperties}
    \centering
    \begin{tabular}{p{0.1\textwidth}p{0.25\textwidth}p{0.35\textwidth}p{0.15\textwidth}}
    \toprule
    \textbf{ID} & \textbf{Informal Description} & \textbf{Structured English Description}& \textbf{Property Pattern and Scope} \\ \midrule
                                \textbf{AV-P01 (a)} & When a failure occurs during a frame's sending, it has to be resolved before the frame is sent completely. & Between Frame\_Generator\_A.start and Frame\_Generator\_A.EOFReset, if Sender\_A.Failure has occurred, then in response Sender\_A.Fail\_Resolved eventually holds. & Response Between
                                  
                                 \\ \midrule
                                 \textbf{AV-P01 (b)} & When a failure occurs during a frame's sending, it has to be resolved before the frame is sent completely. & Between Frame\_Generator\_A.start and Frame\_Generator\_A.EOFReset, if Sender\_A.Failure has occurred, then in response Sender\_A.Resolved eventually holds. & Response Between
                                  
                                 \\ \midrule
                                \textbf{AV-P02} & If one sender detects a collision, then every other simultaneously transmitting sender should detect it before finishing sending. & Between Frame\_Generator\_A.start and Frame\_Generator\_A.EOFReset, if obs.Failure has occurred, then in response obs.Resolved eventually holds. & Response Between \\ \bottomrule
    \end{tabular}
\end{table*}

\subsubsection{Gear Controller}\label{sec:evaluation:gearcontroller}

Lindahl et al.~\cite{LindahlPY01} modeled and verified a gear controller with UPPAAL. The model consists of five automata: a gearbox, an engine, an interface, a clutch, and a controller. The largest automaton, the controller, consists of 24 locations. In total, there are 50 locations and 65 transitions in the model. \citeauthor{LindahlPY01} specified a set of 14 properties that they verified with UPPAAL. 
Unfortunately, it was not possible to semantically map a property pattern to ten out of those 14 properties in the Gear Controller system~\cite{LindahlPY01}. 
Following the results reported by \citet{filipovikj2014reassessing}, this might indicate that the property patterns need extensions, as there appear to be requirements that cannot be expressed with any of the patterns.

From the properties described in Table~\ref{tab:lindahlproperties}, one can notice that from all those properties where the property patterns were applicable, only the time-constrained Response Globally specification pattern was needed (GC-P01 to GC-P04) in the Gear Controller system. Therefore, the matching properties followed the Observer process of our approach.

\begin{table*}[!h]
    \footnotesize
	\caption{Properties of the Gear Controller.}
	\label{tab:lindahlproperties}
    \centering
    \begin{tabular}{p{0.1\textwidth}p{0.25\textwidth}p{0.35\textwidth}p{0.15\textwidth}}
    \toprule
    \textbf{ID} & \textbf{Informal Description} & \textbf{Structured English Description} & \textbf{Property Pattern and Scope} \\ \midrule
                                \textbf{GC-P01} & If the clutch encounters an error while closing, within 200ms the gear control detects this error. & Globally, if Clutch.ErrorClose holds, then in response GearControl.CCloseError eventually holds within 200ms. & Time-constrained Response Globally \\ \midrule

                                \textbf{GC-P02} & If the clutch encounters an error while opening, within 200ms the gear control detects this error. & Globally, if Clutch.ErrorOpen holds, then in response GearControl.COpenError eventually holds within 200ms. & Time-constrained Response Globally \\ \midrule    
                                                         
                                \textbf{GC-P03} & If the gearbox encounters an error while being idle, within 350ms the gear control detects this error. & Globally, if Gearbox.ErrorIdle holds, then in response GearControl.GSetError eventually holds within 350ms. & Time-constrained Response Globally \\ \midrule
                                                             
                                \textbf{GC-P04} & If the gearbox encounters an error while being in neutral gear, within 200ms the gear control detects this error. & Globally, if Gearbox.ErrorNeu holds, then in response GearControl.GNeuError eventually holds within 200ms. & Time-constrained Response Globally \\ \bottomrule
    \end{tabular}
\end{table*}

\subsection{Results}\label{sec:results}

The results of our experiments are reported in Table~\ref{tab:results}. First and foremost, we can notice that all the properties originally specified are satisfied, except for the properties AV-P01(a), AV-P01(b), and AV-P02. These results are exactly as reported by the corresponding papers reporting these system and properties, especially by~\citet{HavelundSLL97}. Due to the failed properties, \citet{HavelundSLL97} re-modeled the Audio/Video system including fixes for the bugs they identified. Applying the fixes in the model proposed by \citet{HavelundSLL97}, we were able to verify the model's correctness through the properties AV-P01'(a), AV-P01'(b), and AV-P02'\footnote{In their original paper, Havelund et al.\@ identified a protocol error and proposed its correction. These renamed properties refer to those originally reported by Havelund et al.}.

\begin{table*}[!t]
	\footnotesize
	\caption{Summary of the results for the BSN system while applying our property-pattern oriented process, where: S+ = states added, T+ = transitions added, UMem = resident memory    to verify the property of interest, Ovh = overhead introduced by our approach compared to the baseline, PTime = time required by our automated flag and/or observer processes accordingly, UTime = time required by UPPAAL to verify the property, Sat? = whether the property was satisfied in UPPAAL. Highlighted are the worst-case scenario for each metric.} 
	\label{tab:results}
	\centering
	\begin{tabular}{cccccccc}
		\toprule
		\textbf{ID} & \textbf{S+} & \textbf{T+} & \textbf{UTime (s)} & \textbf{PTime(s)} & \textbf{UMem(MB)} & \textbf{Ovh} & \textbf{Sat?}\\ \midrule
		\textbf{BSN-P02} & 7 & 16 & 77.3 & 0.660 & 2.6 x10$^{3}$ & 0.90 & yes \\
		\textbf{BSN-P03} & 7 & 0 & \textbf{101.4} & 0.702 & \textbf{3.4 x10$^{3}$} & 1.51 &yes \\
		\textbf{BSN-P04} & 6 & \textbf{18} & 92.9 & 0.707 & 2.6 x10$^{3}$ & 0.87 &yes \\ 
		\textbf{BSN-P05} & 0 & 5 & 172.1 & 0.632 & 2 x10$^{3}$ & 0.42 &yes \\ 
		\textbf{BSN-P06} & 0 & 0 & 51.8 & 0.640 & 1.3 x10$^{3}$ & -0.07 & yes \\ 
		\textbf{BSN-P07} & 0 & 0 & 50,1 & 0.649 & 1.3 x10$^{3}$ & -0.07 &yes \\ 
		\textbf{BSN-P08} & 0 & 0 & 52.4 & 0.653 & 1.3 x10$^{3}$ & -0.07 &yes \\ 
		\textbf{BSN-P09} & 0 & 0 & 55.3 & 0.658 & 1.4 x10$^{3}$ & -0.04 &yes \\ 
		\textbf{BSN-P10} & 0 & 0 & 55.3 & 0.631& 1.4 x10$^{3}$ & -0.03 &yes \\ 
		\textbf{BSN-P11} & 0 & 0 & 57.8 & 0.636 & 1.4 x10$^{3}$ & -0.03 & yes \\ 
		\textbf{BSN-P12} & 0 & 0 & 54.1 & 0.646 & 1.4 x10$^{3}$ & -0.03 & yes \\ 
		\textbf{AV-P01(a)}  & 9  & 16 & 0.7 & \textbf{0.827} & 9.8 & 0.13 & \textbf{no} \\ 
		\textbf{AV-P01(b)}  & 9  & 15 & 0.7 & 0.722 & 9.7 & 0.11 & \textbf{no} \\ 
		\textbf{AV-P02}     & \textbf{10} & 17 & 1.1 & 0.775 & 10.8 &  0.24 & \textbf{no} \\ 
		\textbf{AV-P01'(a)} & 9  & 16 & 6.3 & 0.662 & 91.9 & 9.5 & yes \\ 
		\textbf{AV-P01'(b)} & 9  & 16 & 6.2 & 0.733 & 91.1 & 7.86 & yes \\
		\textbf{AV-P02'}    & \textbf{10} & 17 & 9.6 & 0.677 & 128.5 & \textbf{11.5} & yes \\ 
		\textbf{GC-P01} & 5 & 5 & 0.5 x10$^{-2}$ & 0.618 & 5.9 & -0.15 &yes \\ 
		\textbf{GC-P02} & 5 & 5 & 0.2 x10$^{-2}$ & 0.660 & 6 & 0.05 & yes \\ 
		\textbf{GC-P03} & 5 & 5 & 0.4 x10$^{-2}$& 0.629 & 6.1 & 0.04 & yes \\
		\textbf{GC-P04} & 5 & 5 & 0.3 x10$^{-2}$& 0.641 & 6 & -0.26 & yes \\ \bottomrule
	\end{tabular} 
\end{table*}

From the quantitative perspective, the highest number of locations introduced by our approach was in the UPPAAL model for the Audio/Video system while verifying property AV-P02', i.e., ten (10) locations. Regarding the number of transitions, eighteen more transitions were introduced in the verification of property BSN-P04 in the worst-case scenario. The BSN system model also reached the highest verification time, but affordably below two minutes (BSN-P03). Regarding the time introduced by our tool to apply the Flag and Observer processes, it was quite negligible as it was below one second: 0.827s in the worst-case scenario. Memory consumption to run the BSN in UPPAAL was also the highest among all systems: the upper limit was around 3.4GB for property BSN-P03. However, the property already required a quite high resident memory peak as the overhead introduced by our approach was 1.51 times the original memory consumption. Last but not least, the highest overhead introduced was in the verification of property AV-P02' which was in the order of 11.5 times compared to the baseline system. However, the nominal memory consumption for that property after applying our approach is quite affordable, i.e., 130MB.

Considering the outcomes of our experiments, we come to the conclusion that our approach is validated since all the properties originally reported in our experiments were satisfied after applying our approach. Therefore, the experiments show that we have preserved the original behavior of the UPPAAL systems. In addition, the computational feasibility of our approach has also presented itself as quite affordable by the metrics extracted from running our experiments and reported in Table~\ref{tab:results}.

\subsection{Threats to Validity}

Given the nature of our experiments, some threats could have compromised the validity of our results. In the first place, two out of the three systems upon which we conducted our experiments did not have their original UPPAAL models available. While~\citet{RodriguesCR0P18} already provided the UPPAAL models and properties for the BSN, the other two systems in our experimentation~\cite{LindahlPY01,HavelundSLL97} did not provide a repository for the original UPPAAL models. Nevertheless, it was possible to reproduce all those examples in UPPAAL and check whether our approach could handle their property requirements. In this case, we had to re-build the UPPAAL models based on the models' descriptions provided in their corresponding manuscripts, as previously reported. To overcome the threat, we made it evident that not only the models were correctly modeled and executed in UPPAAL, following the expected behavior, but also that the same original system properties were satisfied. 
Given that we reached the same results as those reported in the original paper, we argue that this threat has been properly dealt with. The exception, however, was in the Gear Controller system, where we could not replicate all the reported results. We were not able to replicate properties (1) and (2) of the original Gear Controller system, as it was not possible to identify in their work how non-atomic propositions should be handled with the time-annotated leads-to operator introduced in their paper.
 
Concerning the ability to draw the correct conclusion from our results, it relies on the correct mapping between the systems' properties originally reported and the property pattern we apply. First and foremost, prior solid knowledge of property patterns must be assured. We make that evident, as the appropriate and sound mapping between the property specification pattern and its corresponding observer automata has been extensively reported in our comprehensive PSP catalog. Then, to go through the corresponding observer for each system property of our experiments, we went through a peer-review process after a careful analysis of (i) the informal description provided in the original paper, (ii) the formal description of the original properties in TCTL, and (iii) the reproduction of the UPPAAL models. Once there was a clear understanding of the systems' behavior and their specified properties, it was straightforward to map them to the corresponding property pattern. In addition, for each analyzed property there is its corresponding UPPAAL model, which makes clear that each property is individually analyzed and its results are individually reported by the model checking tool.

Regarding the reliability of our measures, we have reported the UPPAAL models, properties, and results of our experiments in the public repository. All this information can be reproduced, following the runtime environment described in Section~\ref{sec:setup}. Moreover, the consistency of the results whether the verified properties are satisfied is assured by the maturity of the UPPAAL model checker.

Regarding our approach and experiments, we are aware of the following limitations.
Our experiments are limited to the three UPPAAL systems reported and we cannot generalize our results. 
A limitation of our approach is that properties can only refer to traceable states in terms of locations in the system model. When this limitation became apparent in one of our experiments, we had to manually refactor the UPPAAL model and property to be analyzed to make the application of our approach feasible. Although we have managed to overcome this limitation in our experiments, there may be cases where this refactoring procedure may be hindered. 
Another limitation of our approach is that we can only express properties that follow a pattern from our catalog. On the one hand, we rely on a comprehensive catalog collected by \citet{AutiliGLPT15} that is based on the catalogs by \citet{DwyerAC99} and \citet{KonradCheng05} for qualitative respectively real-time requirements. On the other hand, we still encountered some properties that cannot be mapped to patterns of our catalog (\textit{cf.} Gear Controller in Section~\ref{sec:evaluation:gearcontroller}). This observation is in line with the results by \citet{filipovikj2014reassessing}, who investigated how sufficient the patterns by \citet{DwyerAC99} were to express requirements gathered from industry. They reported that most but not all of them were expressible.
Likewise, \citet{Post12MHP} conducted a case study in the automotive domain at Bosch and reported that the majority of requirements but not all of them could be reformulated using the patterns by \citet{KonradCheng05}. They further state that the sufficiency of such a pattern catalog cannot be proven.
Therefore, not only the comprehensiveness of a pattern catalog but also its ability to reformulate all sorts of requirements is still an open question in the literature and requires further investigation to either refute or confirm the comprehensiveness of such a catalog. Naturally, the same holds for the catalog we propose. However, the purpose of the presented work is not to analyze the sufficiency of an existing catalog and to extend the catalog if needed. In contrast, our goal was to leverage existing patterns in UPPAAL, thereby relying on the most comprehensive catalog~\cite{AutiliGLPT15}.

Finally, additional scalability studies should be performed to investigate the space/time overhead introduced during the model augmentation phase for a more thorough conclusion.

\section{Related Work}\label{sec:Related}

In this section, we discuss related work according to the topics covered in this paper. The first one is about theoretical foundations for observer-based model checking. The second topic concerns practical work that provides generative observer-based approaches where transformation processes were applied. In the third topic, we present related work on model checking with specification patterns.

\subsection{Theoretical Observer-based Approaches}

The approach of observer timed automata to real-time system verification might have been first suggested in the literature by Havelund et al.~\cite{HavelundSLL97}. This approach has been used to model check in practice real-life systems such as the B\&O power controller and some timed safety instrumented systems~\cite{LahtinenVBFNH12}. In~\cite{HavelundLS99}, Havelund et al. apply the three techniques, i.e., \@ flag, debt, and observer, to annotate their system model by adding new variables or communication actions, and then observe these. They do this either by mentioning the variables in the formulae to be verified (the first two techniques) or by letting the new communication actions synchronize with a furthermore added observer automaton (the third technique). Case studies thereof indicate that the approach is effective.

Further on, observer automata were used by Aceto et al.~\cite{AcetoBBL03} for model checking temporal properties specified in safety model property language (SBBL) on timed automata. Their automata construction encodes a temporal logic formula. The approaches by Gerth et al.~\cite{GerthPVW95} and Tripakis et al.~\cite{Tripakis09} perform LTL model checking on timed Büchi automata and encode the properties as temporal logic formula in automata as well. However, all those approaches also come with quite a few limitations such as the manual construction of those automata as well as the error-proneness process to adapt and observe the original system model. Moreover, the comprehensiveness of the observers represented is not assured as it is in our work in the form of the pattern specification catalog.

\subsection{Generative Observer-based Approaches}

Jensen et al.~\cite{JensenLS00} present a method for scaling up the real-time verification tool UPPAAL by complementing it with methods for abstraction and compositionality through the notion of timed ready simulation. Then, they proposed a method for automatically testing for the existence of a timed ready simulation between real-time systems using the UPPAAL tool. Later on, Heinzemann et al.~\cite{HeinzemannBDS15} extend the compositional verification approach of Jensen et al.~\cite{JensenLS00} in particular by using different kinds of refinement definitions including an automatic selection of the most suitable refinement definitions to construct test automata for refinement checking.
 
Li et al.~\cite{LiBDLNP10} use Live Sequence Chart (LSC), a message-only untimed chart of real-time systems for property specification and system modeling based on UPPAAL. In their work, Li et al. specify safety and liveness properties for timed automata as LSC. They translate their LSC into a timed observer automaton that reaches an error location if the property is violated, following the same principle as we apply for the safety check of our observers. Therefore, that work requires the introduction of another layer of the system model as LSCs following a scenario-based verification approach. 

Andr\'{e}~\cite{Andre13} proposes a set of correctness patterns encoding common properties met when verifying concurrent real-time systems. He identifies commonly used properties of correctness for real-time systems. Then, he proposes an abstract syntax for each pattern, followed by the translation of each pattern to an observer instantiated in both timed automata and stateful timed CSP. Finally, he provides a concrete syntax for the patterns, implemented in the IMITATOR tool~\cite{AndreFKS12}. 
Later on, Andr\'{e} and Petrucci~\cite{AndreP15} propose a set of patterns that encode common specification or verification components when dealing with concurrent real-time systems. There, they provide formal semantics for those patterns, as time Petri nets, and show that they can encode previous approaches~\cite{Andre13}. Besides the limitation of their set of supported property patterns to their stateful timed CSP language~\cite{SunLSLSA13}, their work, like the work of \citet{Gruhn06L}, might not be directly applicable to existing standard model checking tools for real-time systems such as \mbox{UPPAAL}. 

\citet{BrabermanKOVSC_TSE} proposed the Visual Time Event Scenarios (VTS), a visual formalism to express and model check complex event-based requirements for real-time systems. A tool was also developed to translate visually specified scenarios into observer timed automata to check satisfaction of the stated scenarios. Further on, \citet{AsteasuainB17} propose the omega-feather weight visual scenarios ($\omega$-FVS), which is a declarative language founded on graphical scenarios and capable of expressing $\omega$-regular properties. As in \cite{BrabermanKOVSC_TSE}, one can use $\omega$-FVS to  automate the generation of rule scenarios (anti-scenarios) that show how things could go wrong and violate the rule at stake. Indeed, their work could be an alternative to express property specifications in the form of visual specifications instead of the temporal logic formalism. However, the expressiveness of their formalism and the practical relevance of the approach in terms of the PSP catalog are not clear.

\subsection{Specification Pattern-based Approaches}

Post et al. have proposed an automated analysis of formal requirements~\cite{PostH12} for properties such as consistency and vacuity~\cite{Post12MHP}. To do so, requirements must have been already formalized or mathematically described. More recently, Post et al. have proposed the Req2Spec method~\cite{Req2Spec:REFSQ22} where they integrate their previous work~\cite{Post12MHP} into HANFOR~\cite{Hanfor:2021}---an industry scale tool based on the specification pattern system by \citet{KonradCheng05}.  By these means, they ``automatically translate the formal specifications into logics for downstream processing''~\cite{Req2Spec:REFSQ22}. Although our work aims at a broader scope, possible extensions of our pattern catalog could target their set of properties as well.

Another recent approach to formalize and analyze requirements is \emph{FRET}, introduced by \citet{fret} and building upon the specification patterns by \citet{DwyerAC99}. One of the core improvements of FRET is the support of both future-time as well as past-time temporal logic.

\citet{Rebeca:SEFM20} propose a formal verification process where structured requirements in the form of GIVEN-WHEN-THEN are encoded into behavior models (state diagrams and sequence diagrams) from which their Rebeca model (an actor-based language) is derived and checked against specified (safety) properties. We believe their work could be potentially leveraged by targeting the structured requirements of the PSP catalog, which could provide a more comprehensive set of properties they could verify. Moreover, model checking UPPAAL models, as opposed to Rebeca models, provides more comprehensive analyses other those safety requirements.

\subsection{Summary}
In our work, we have focused on a more fine-grained way to have the observer interact with the system, following the solid foundations provided by \citet{HavelundLS99} and without introducing further modeling overhead to the existing system. In all those related work, the fact that the properties of interest might not be directly verifiable in a widely-used tool for real-time systems verification like UPPAAL requires further strategies to be able to explore the benefits of an observer-based verification process and the beneficial impacts of using property specification patterns. For instance, such works do not provide ways to automatically adjust the model so it can be rendered automatically observable as we do in our approach. Without such an automated step, the applicability of an observer-based approach may be hindered as it is still subject to further human efforts to adapt the system model. Moreover, our work relies on the solid and comprehensive foundation of property specification patterns~\cite{AutiliGLPT15,KonradCheng05,DwyerAC99}. In this way, we are able to make evident not only the practical relevance of our work, but also to automatically build an observer for each property pattern without changing the semantics of the model  and provide a seamless reachability checking for real-time systems.
\section{Conclusion and Future Work}\label{sec:Conclusion}

Property specification patterns aim for bridging the gap between practitioners and model checking. As typical for patterns, the property specification patterns are organized in a catalog, which comprises the ``best practices'' in system specification and represents an attempt to capture proven solutions in a single framework. 
However, there exists a gap between the property specification patterns~\cite{AutiliGLPT15} which use the full expressive power of temporal logic formalisms and UPPAAL which supports only a subset of such formalisms.
Thus, UPPAAL does not support the majority of the property specification patterns as the corresponding TCTL formulae cannot be expressed in UPPAAL.
Moreover, existing work does not bridge the gap between \textit{all} property specification patterns and existing model checkers, even beyond UPPAAL, considering qualitative and real-time requirements by leveraging an automated approach to generate observers. 

The contribution of this work is that we have closed this gap by developing a comprehensive property specification pattern catalog for UPPAAL. The catalog supports qualitative and real-time properties that are specified using patterns and automatically mapped to UPPAAL for verification. To achieve this mapping, we have concretized and automated the manual flag, debt, and observer techniques proposed by \citet{HavelundLS99} for \textit{all} qualitative and real-time patterns of the catalog by \citet{AutiliGLPT15}, except of the real-time variants of the \textit{Precedence} patterns that are not feasible with UPPAAL. Especially, we have specified UPPAAL-compatible formula and observer templates that are specific for each pattern and that are collected in our publicly available catalog. Moreover, we have developed an automated generator to produce concrete formulae and observers based on the templates given a pattern-based property specification from a user. The resulting formulae and observers can be directly used for verification in UPPAAL. Thus, we have leveraged the benefits of property specification patterns to the widely-used UPPAAL by enabling practitioners to specify properties in a pattern-based way, that is, without using a temporal logic, \textit{and} to verify these properties in UPPAAL. This approach promises to ease the use of model checking with UPPAAL in practice.

We have evaluated our catalog on three real-time systems from literature~\cite{LindahlPY01,HavelundSLL97, RodriguesCR0P18}, two of which have been applied in an industrial setting. These systems have been originally verified with UPPAAL using manually created observer automata. Applying our catalog to these systems, we demonstrate that our catalog can express the properties of interest and automatically generate corresponding observer automata and formulae. Using these generated observer automata and formulae, we were able to reproduce the verification results reported in literature for these systems, which provides evidence for the validity of our catalog.

As future work, we envision to prove by (weak) bisimulation that the annotations we make to the system model do not alter its specified behavior. We also plan to prove that the observers that we build are correct with respect to the TCTL formulae in the catalog by \citet{AutiliGLPT15}. Last, but not least, we plan to apply our approach to observers/monitors at runtime.

\section*{Acknowledgments}
This research work is part of the Safe.Spec project funded by the German Federal Ministry of Education and Research (01IS16027). This study was financed in part by the Coordenação de Aperfeiçoamento de Pessoal de Nível Superior – Brasil (CAPES) – Finance Code 001 and the Alexander von Humboldt Foundation.\par

%\bibliographystyle{elsarticle-num-names}
%\bibliography{bibliography.bib}

\end{document}